\documentclass[twocolumn]{aastex63}

\usepackage{graphicx}
\usepackage{dcolumn}
\usepackage{bm}
\usepackage{ulem}

\begin{document}

\shorttitle{The Astrometric Gravitational Wave Background}
\shortauthors{Darling}

\title{A New Approach to the Low Frequency Stochastic Gravitational Wave Background:  Constraints from Quasars and the Astrometric Hellings-Downs Curve}

\author{Jeremy Darling}
 \affiliation{Center for Astrophysics and Space Astronomy \\
Department of Astrophysical and Planetary Sciences \\
University of Colorado, 389 UCB \\
Boulder, CO 80309-0389, USA}
\email{jeremy.darling@colorado.edu}


\begin{abstract}
  We present new astrometric constraints on the stochastic gravitational wave background and construct
  the first astrometric Hellings-Downs curve using quasar proper motions.
  From quadrupolar vector spherical harmonic fits to the Gaia proper motions of 1,108,858 quasars,
  we obtain a frequency-integrated upper limit on the gravitational wave energy density, $h_{70}^2\Omega_{GW} \leq 0.023$
  (95\% confidence limit), for frequencies between 11.2 nHz and $3.1\times10^{-9}$ nHz ($1.33/t_0$).  
  However, from the astrometric Hellings-Downs curve that describes the correlated proper motions between
  2,104,609,881 quasar pairs as a function of their angular separation, we find a stronger constraint: 
  a characteristic strain of $h_{c} \leq 2.7 \times 10^{-12}$ for $f_{\rm ref} = 1$~yr$^{-1}$ and
  $h_{70}^2\Omega_{\rm GW} \leq 0.0096$ at 95\% confidence.
  We probe down to $\pm$0.005 $\mu$as$^2$ yr$^{-2}$ in correlated power and obtain the lowest astrometric limit to date.  This is also the first time that optical wavelength astrometry surpasses limits from radio-frequency interferometry.
This astrometric analysis does not yet reach the sensitivity needed to detect the pulsar timing-based red gravitational wave spectrum extrapolated to the quasar gravitational wave sensitivity window, assuming that the turnover in the spectrum occurs at $\sim$1 nHz for massive black hole binaries.  The limits presented here may exclude some exotic interpretations of the stochastic gravitational wave background.
\end{abstract}

\section{\label{sec:intro}Introduction}

Gravitational waves have now been detected from binary black hole mergers at frequencies of 35--250~Hz by LIGO and as a gravitational wave background via 
pulsar timing at 2--28 nHz \citep{abbott2016,agazie2023}.  The  gravitational waves seen in pulsar timing show a red spectrum that is consistent with a stochastic background
rather than from single sources, but the provenance of the background remains unknown.  The most likely source
is massive black hole binaries, but other sources such as a cosmological primordial background are possible \citep{afzal2023}.  If the pulsar timing
signal arises from massive black hole binaries, then it will have a red spectrum with the characteristic dimensionless
strain $h_c \propto f^{-2/3}$ but should turn over at some frequency, most likely $\lesssim 1$~nHz \citep[e.g.,][]{sesana2004,enoki2007}.

The steep spectrum presents an opportunity for astrometric gravitational wave detection methods, which span lower frequencies but
are typically less sensitive to strain than pulsar timing.  Astrometric methods, whether they detect gravitational waves or produce an upper limit, can inform
the interpretation of the pulsar timing result, can resolve the ambiguity of the source of the signal, and could potentially
inform our understanding of the eccentricity of massive black hole binaries and their environmental coupling by probing
the low-frequency turnover in the gravitational wave spectrum  \citep[e.g.,][]{chen2017,kelley2017}.  

Pulsar timing residuals measure the correlated radial motions induced by a stochastic
gravitational wave background that follow the Hellings-Downs curve \citep[HD curve;][]{hellings1983}, but gravitational waves produce correlated transverse angular motions as well \citep[e.g.,][]{pyne1996,gwinn1997,book2011,darling2018}.
Reductively, gravitational waves have a characteristic strain that manifests in pulsar timing as
$h_c \sim \Delta f_s / f_s$  ($f_s$ is the pulsar spin frequency) while angular deflections scale as $h_c\sim \delta\theta$.
In practice, angular deflections are measured as proper motions $\mu$ over a timescale $\Delta t = 1/f$:  $h_c \sim \delta\theta \sim \mu \Delta t \sim \mu/f$.  Pulsar timing or proper motion correlations (or cross-correlations between the two) will therefore measure $h_c^2$, which is
proportional to the gravitational wave energy density \citep[e.g.,][]{moore2015}.

Angular motions can be described using vector spherical
harmonics \citep[VSH;][]{pyne1996,gwinn1997,book2011,darling2018} or using generalized HD curves that include
astrometric correlations \citep[e.g.,][]{mihaylov2018,obeirne2018,qin2019,caliskan2024}.  VSH-based analyses of quasar proper motions have used Very Long Baseline Interferometry (VLBI) radio observations \citep{gwinn1997,titov2011,darling2018,jaraba2023} and Gaia\footnote{\citet{gaia2016}} optical proper motions \citep{jaraba2023}.  No gravitational wave signature has been detected to date via astrometry, and the best limits on the frequency-integrated energy density were obtained by \citet{jaraba2023}:
$h^2_{70}\Omega_{GW} \leq 0.087$ for frequencies in the range $4.2\times10^{-18}$ Hz to $1.1\times10^{-8}$ Hz based on
Gaia proper motions \citep{gaia2023} and 
$h^2_{70}\Omega_{GW} \leq 0.024$ for $5.8\times10^{-18}$ Hz to $1.4\times10^{-9}$ Hz using the VLBI catalog presented in
\citet{truebenbach2017} and analyzed by \citet{darling2018}.

The dominant VSH mode produced by a stochastic gravitational wave background
is quadrupolar ($\ell = 2$).  However, the Gaia quasar proper motions show VSH power at all scales, starting with E- and B-mode dipoles
($\ell \geq 1$; \citealt{klioner2022} and this work).  
Aside from the E-mode dipole associated with the secular aberration drift caused primarily by the barycenter acceleration about
the Galactic Center \citep{titov2011,truebenbach2017,GaiaAccel2021}, the significant VSH modes are most likely systematic to
the Gaia observations and astrometric solutions \citep{klioner2022}.
This issue with Gaia effectively sets a floor on the stochastic gravitational wave background that can be detected using VSH fitting.  

Here we present the first astrometric HD analysis of Gaia quasar proper motions and show that this technique does not
seem to find a systematic floor, offering unprecedented astrometric sensitivity to gravitational waves.  
In Section \ref{sec:theory} we review the astrometric signal expected from a stochastic gravitational wave background.
We describe the data sources and treatment in Section \ref{sec:data} and the astrometric methods in 
Section \ref{sec:methods}.  Section \ref{sec:results} presents new limits on the background as measured by VSH and HD methods and examines the error budgets, systematic effects, and possibly intrinsic scatter in the astrometric correlations.
Section \ref{sec:discussion} assesses the implications of these results in light of the pulsar timing red spectrum
and discusses future work. 

We assume a flat $\Lambda$CDM cosmology with $H_0  =70$ km s$^{-1}$ Mpc$^{-1}$ and $\Omega_{M,0} = 0.3$.  This is used to calculate lookback times, the corresponding gravitational wave frequencies, and the cosmological
gravitational wave energy density.  It is worth noting that the Hubble
constant is a rate, which can be written as an angular rate for comparison to extragalactic proper motions:  $H_0 = 15$~$\mu$as~yr$^{-1}$.

\section{\label{sec:theory} Theory}

\begin{figure}[t]
\begin{centering}
  \includegraphics[scale=0.5,trim= 30 30 30 0,clip=false]{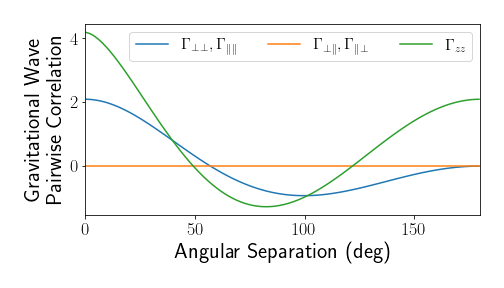}
  \caption{Generalized Hellings-Downs curves showing the predicted pairwise correlations of radial and angular motions produced by 
    transverse traceless stochastic gravitational waves as a function of angular separation.   The blue track shows the
    correlated angular motions along and perpendicular to the great circles connecting pairs of objects
    ($\Gamma_{\parallel\parallel}$ and
    $\Gamma_{\perp\perp}$, respectively).  The orange track depicts the angular cross-correlations, $\Gamma_{\perp\parallel}$ and $\Gamma_{\parallel\perp}$, which are expected to be null-valued for all angular separations.
    The green line shows the canonical Hellings-Downs curve seen in pulsar timing that is caused by radial motions
    ($\Gamma_{zz}$).  }
\label{fig:theory}
\end{centering}
\end{figure}

The effect of gravitational waves on proper motions can be described in two mathematically equivalent ways:  using vector spherical harmonics \citep[VSH;][]{mignard2012} or angular correlation functions \citep[e.g.,][]{mihaylov2018,obeirne2018,qin2019,caliskan2024}.  The VSH
manifestation of gravitational wave proper motions is in $\ell\geq2$ modes with equal power in
spheroidal ``E'' modes and toroidal ``B''  modes and is dominated by the quadrupole \citep{pyne1996,gwinn1997,book2011,darling2018}.
The total quadrupolar VSH power $P_2$ can be related to the cosmological gravitational wave energy
density $\Omega_{\rm GW}$ as 
\begin{equation}\label{eqn:Omega}
    \Omega_{\rm GW} = {6\over5}\,{1\over4\pi}\,{P_2\over H_0^2} 
                              = 0.00042\, {P_2\over (1\ \mu\rm{as\ yr}^{-1})^2}\,h_{70}^{-2}
\end{equation}
where the factor of $6/5$ corrects for the $\ell >2$ modes \citep[e.g.,][]{pyne1996,gwinn1997,book2011}.
The details of the VSH signal and fitting are well-described elsewhere \citep[e.g.,][]{mignard2012,darling2018}, but
we summarize our specific methods in Section \ref{subsec:VSH}.  

Like the VSH, angular proper motion correlations will have two equal-power (and identical)
terms.  For the astrometric HD curves, the two terms correspond to proper motions along ($\parallel$) and across ($\perp$) the great circle connecting pairs of objects, which could be stars, pulsars, quasars, or other luminous sources \citep{book2011,mihaylov2018,obeirne2018,qin2019,golat2022,caliskan2024,inomata2024}.  The angular correlations of perpendicular proper motions $\Gamma_{\perp\perp}(\theta)$  and parallel proper motions $\Gamma_{\parallel\parallel}(\theta)$ depend on the separation $\theta$ between pairs of objects as 
\begin{eqnarray}\label{eqn:gamma}
  \Gamma_{\perp\perp}(\theta) = \Gamma_{\parallel\parallel}(\theta) =  {2\pi\over3} \left(1-7\sin^2\left(\theta\over2\right)\right. \nonumber \hspace{40pt}\\
  \left.-12\sin^2\left(\theta\over2\right) \tan^2\left(\theta\over2\right) \ln\left[\sin\left(\theta\over2\right)\right]\right).
\end{eqnarray}
This is the astrometric version of the radial HD curve used in pulsar timing (and depicted in Figure \ref{fig:theory} as $\Gamma_{zz}(\theta)$ for reference).  
For a transverse traceless isotropic unpolarized gravitational wave background, the cross-correlation terms vanish:
$\Gamma_{\perp\parallel}(\theta) = \Gamma_{\parallel\perp}(\theta) = 0$ for all $\theta$.  This provides a powerful check on observational systematics
because all four angular correlation functions can be calculated from 2D (vector) proper motions:
the null-valued correlations provide a direct empirical measurement of any systematics present in the observations
(see Section \ref{sec:methods}).
Figure \ref{fig:theory} shows the expected proper motion angular correlations.  

The angular correlation between vector proper motions $\bm\mu$ is the observable quantity; it is the amplitude of this angular correlation signal that this analysis will seek.  The average product of projected
proper motions over all pairs of objects $(n,m)$ with separation $\theta$ is
\begin{equation}\label{eqn:CtoIGamma}
    C_{ij}(\theta)  \equiv  \left\langle \mu_{n,i}\ \mu_{m,j} \right\rangle_\theta
      = f^3 I_{\rm GW}(f) \Gamma_{ij}(\theta)
\end{equation}
where $i$ and $j\ \epsilon\ \{\parallel, \perp\}$ and $I_{\rm GW}(f)$ is the gravitational wave intensity, which is related to the
characteristic strain as
\begin{equation}
  h_c^2(f) = 16 \pi f I_{\rm GW}(f)
\end{equation}
\citep{caliskan2024}.
The cosmological energy density is related to the characteristic strain as
\begin{equation}
  H_0^2 \Omega_{\rm GW}(f) = {2\pi^2\over 3} f^2 h_c^2(f)
\end{equation}
\citep{moore2015}.
Because $\Gamma_{ij}(\theta) =0$ for $i\neq j$ for a transverse traceless unpolarized gravitational wave background, 
the proper motion correlations provide a measurement of the characteristic strain for $i=j$:
\begin{equation}
  h_c^2(f) = 16 \pi {C_{ii} (\theta) \over f^2 \Gamma_{ii}(\theta)}
\end{equation}
and the gravitational wave energy density:
\begin{equation}
  H_0^2 \Omega_{\rm GW}(f) = {32\pi^3\over 3}  {C_{ii} (\theta) \over \Gamma_{ii}(\theta)}.
\end{equation}
Expressing $H_0$ as an angular rate, we obtain the HD analog of Equation \ref{eqn:Omega} that connects
the observable quantity to the gravitational wave energy density:
\begin{equation}\label{eqn:OmegaHD}
  \Omega_{\rm GW}(f) = {1.5 \over \Gamma_{ii}(\theta)} {C_{ii} (\theta) \over (1\ \mu{\rm as}\ \rm{yr}^{-1})^2} h_{70}^{-2}.
\end{equation}
Section \ref{subsec:HD} describes the process for projecting equatorial coordinate proper motion vectors parallel
and perpendicular to great circles connecting pairs of objects in order to calculate $C_{\perp\perp}(\theta)$ and
$C_{\parallel\parallel}(\theta)$.


\section{\label{sec:data} Data}

For this study, we select the Quaia quasar catalog, a compilation of Gaia-detected quasars
in Gaia DR3 \citep{gaia2023,quaia}.  This sample has a large sky footprint, covering
73\% of the sky nearly uniformly outside the $A_V > 0.5$ mag Galactic Plane region, which makes
this sample superior for all-sky measurements of proper motion fields.  The catalog redshifts are not
essential for this work because the proper motion signals are distance-independent.  We do, however,
use the redshift distributions to estimate the lower bound on gravitational wave frequency sensitivity
(Section \ref{sec:methods}).  

The Quaia samples include a 755,850-object catalog with Gaia $G<20.0$ and a 1,295,502-object catalog with $G< 20.5$ \citep{quaia}.
We select Quaia quasars that have Gaia DR3 ``\texttt{5p}'' astrometric solutions:  this reduces the samples to 688,367 quasars with $G<20.0$ and 1,111,025 quasars with $G<20.5$.
In this work, we focus on the larger catalog because, while the
magnitude cut does impact the Gaia astrometric solutions, there are sources in the larger catalog with low-uncertainty proper motion measurements.
Given the median proper motion uncertainty of the sample, $\sim$200~$\mu$as~yr$^{-1}$,
we do not expect that any individual quasar will show significant real proper motion in the Gaia DR3.  Those
that do are spurious and are removed from the sample as described below.


The Gaia DR3 duration is 34 months, corresponding to a gravitational wave frequency of
$f_{\rm GW} = 11.2$ nHz.  Because quasar proper motions are secular measurements, as opposed to pulsar timing
residuals, this is the upper limit on the frequency range spanned by the astrometry.  The
lower limit is set by the lookback time of the quasars, and is of order $3\times10^{-18}$ Hz (Section \ref{sec:methods}).
The gravitational wave frequencies probed by this method therefore span more than 10 orders of magnitude.
Note that proper motions are quoted as the angular motion per year but the actual frequency $1/f = 1$~yr is not
directly sampled.

The DR3 quasar sample used for the Gaia astrometric solution, the third Gaia Celestial Reference Frame (Gaia-CRF3), shows correlations in proper motions on many angular scales, both toroidal and spheroidal \citep{klioner2022}.
The secular aberration drift, the apparent
E-mode dipole motion of quasars toward the Galactic Center caused by the Solar acceleration
\citep{titov2011,truebenbach2017,GaiaAccel2021},
is presumably the only physical signal in these correlations; all other vector spherical harmonic E-modes and B-modes are likely systematic to the observations
and/or the astrometric solutions.  It is noteworthy that a significant B-mode dipole exists in the Gaia DR3 quasars
(Section \ref{subsec:VSH}), which indicates a
rotating reference frame (or a rotating universe!).  

The question at hand is whether the Gaia quasars' systematic global correlated proper motions prevent improved constraints
on the stochastic gravitational wave background.  As demonstrated below, VSH fitting finds a non-physical systematic floor
in the quadrupolar power associated with gravitational waves.  But do pairwise astrometric correlations show the same
effect?


\section{\label{sec:methods} Astrometric Methods}

In order to compare the two astrometric gravitational wave detection methods,
we present both VSH and astrometric HD fits to subsets of the Quaia $G<20.5$ \texttt{5p} catalog.  The former shows significant dipole and quadrupole
spheroidal and toroidal power, which motivates the astrometric HD analysis as a means to examine and
perhaps overcome systematics in the Gaia DR3 proper motion solutions.

\subsection{Vector Spherical Harmonics}\label{subsec:VSH}

While Gaia-based quasar proper motion samples that overlap the Quaia catalog have been used for VSH analysis focused on gravitational waves, such as the \citet{jaraba2023} study using 773,471 quasars from Gaia DR3, the full Quaia catalog has not been analyzed.  For completeness and as a comparison to the HD analysis, we present VSH dipole and quadrupole fitting here.

We employ the Quaia $G<20.5$ catalog, excluding highly significant proper motions by imposing a signal-to-noise $< 4$ cut in the proper motion amplitude.  The resulting sample contains 1,108,858 quasars (99.8\% of the \texttt{5p} sample).  In this sample, the redshift quartiles are $z=1.07$, 1.51, and 2.01, corresponding to lookback times
of 8.0, 9.3, and 10.3 Gyr and minimum GW frequencies of $4.0\times10^{-18}$, $3.4\times10^{-18}$, and
$3.1\times10^{-18}$ Hz.  Following previous work, we use the 75th redshift percentile to set the lower bound on the
effective integrated frequency \citep{darling2018,jaraba2023}.  For the adopted cosmology, this corresponds to
$1.33/t_0$.

We fit and remove the E- and B-mode dipoles (E1 and B1) from the proper motion data using a
simple least-squares optimization.
Simultaneous fits for the dipoles and quadrupoles are not significantly different from a sequential fit-and-subtract
method with increasing $\ell$, but we do the fitting of $\ell$ modes sequentially in order to match the dipole subtraction process used for the astrometric HD analysis below.

The best-fit E-mode dipole shows a streaming motion of $5.20\pm0.31$~$\mu$as~yr$^{-1}$ (Table \ref{tab:vsh}).  
The E1 apex lies at the equatorial coordinates $273\fdg8(3\fdg7)$ $-15\fdg8(3\fdg5)$, somewhat offset from the
expected direction of the secular aberration drift, which is toward the Galactic Center.  The aberration drift is the result of
the solar system barycenter acceleration toward the Galactic Center as the Sun orbits the Galaxy.  The changing
aberration of distant quasars causes them to appear to move toward the Galactic Center \citep{titov2011,truebenbach2017,GaiaAccel2021}.

The best-fit B-mode dipole has a significant amplitude of $3.12\pm0.32$~$\mu$as~yr$^{-1}$ that can be thought of as the
angular velocity of rotation, either of the observer or the reference frame (the universe at large).  
The B1 apex\footnote{By ``apex'' we mean the center of an anticyclonic flow in the southern hemisphere, which
  follows the right-hand rule.} lies at $27\fdg9(6\fdg2)$ $-26\fdg9(6\fdg2)$, equatorial.  This is unlikely to be physical under the assumption that the Gaia reference frame shows rotation.  Regardless, we subtract both of the highly significant dipole signals from the Quaia proper motions before fitting quadrupoles to the proper motion vector field.  

The two quadrupolar VSH modes have five coefficients each, and their sum in quadrature is (modulo factors of two) the power in each mode.  Appendix \ref{sec:vsheqns} details of the functional forms fit to the proper motion data, the computation of mode power, and the expressions for dipole amplitude and direction.  One can employ more sophisticated VSH fitting that de-weights significant outlier data or that
employs MCMC methods as described in \citet{darling2018}, but
the differences in outcomes are negligible given the size of the sample and the significance of the power
in the dipole and quadrupole modes.

\subsection{Astrometric Hellings-Downs Correlations}\label{subsec:HD}

For this treatment, we select all quasars with proper motion amplitude less than 100~$\mu$as~yr$^{-1}$, which reduces the
sample to 64,879 quasars.  The reason for the down-select to this best-measured subset is that the error-weighted
two-point proper motion correlations described below show diminishing uncertainty improvement as the proper motion
cut is increased, reaching a minimum around a cut of 100~$\mu$as~yr$^{-1}$.  For example, selection of proper motion
amplitude $<110$~$\mu$as~yr$^{-1}$
produces a sample of 76,588 quasars but $\sim$9\% larger median uncertainties in the astrometric HD curve.

\begin{figure}[t]
\begin{centering}
  \includegraphics[scale=0.25,trim= 20 20 20 20,clip=false]{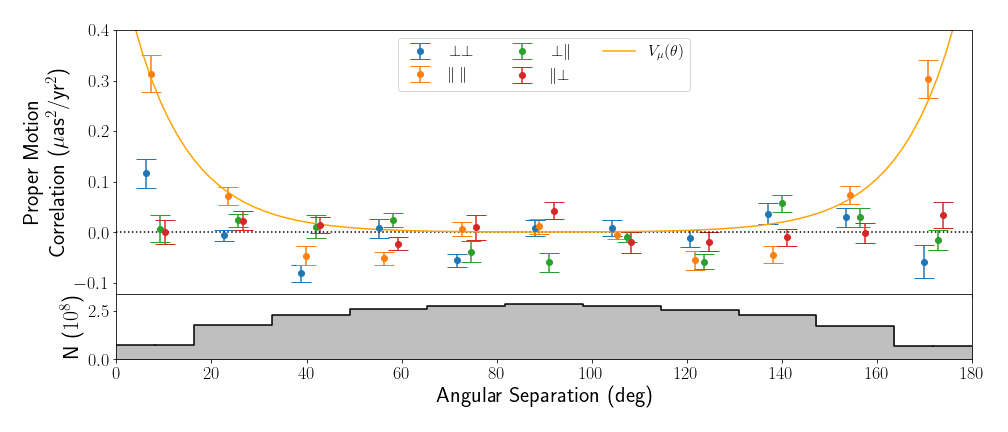}
  \caption{Proper motion power in the four correlations
    (Equations \ref{eqn:Cperpperp}--\ref{eqn:Cparperp}) vs.\ the angular separation of
    quasar pairs for a sample of 67,917 quasars.  These were selected based on a proper motion amplitude
    uncertainty $< 150$~$\mu$as~yr$^{-1}$ and show a systematic in the $C_{\parallel\parallel}$ correlation
    (Equation \ref{eqn:Cparpar}).
    The orange line shows the best fit of this systematic, described by $V_\mu(\theta$) in
    Equation \ref{eqn:sysfit}, a modified version of the quasar proper
    motion uncertainty correlation found in Gaia eDR3 by \citet{lindegren2021}.
  The points are laterally offset a few degrees from the center of each angular separation bin for clarity.  
   The lower histogram shows the distribution of quasar pairs, scaled by $10^8$.}
\label{fig:PMerrSelection}
\end{centering}
\end{figure}

We also explored selection based on proper motion amplitude uncertainty. Choosing uncertainties $<150$~$\mu$as~yr$^{-1}$
produces a similar-sized sample of 67,917 quasars, but this sample shows a notable systematic correlation resembling
the proper motion uncertainty correlation found by \citet{lindegren2021} in Gaia eDR3 quasars.
At small and large angular separations,  proper motions parallel to the great circles connecting pairs of quasars
are significantly correlated, as demonstrated in Figure \ref{fig:PMerrSelection}
(the calculation and meaning of the proper motion correlations are described below).
\citet{lindegren2021} fit the small-separation systematic using an exponential with characteristic scale of $12^\circ$.  We
fit the systematic plotted in Figure \ref{fig:PMerrSelection} using exponentials at high and low angular separations using the same angular scale:
\begin{equation} \label{eqn:sysfit}
  V_\mu(\theta) = A_{\rm sys} \left( e^{-\theta/12^\circ} + e^{(\theta-180^\circ)/12^\circ}\right)
\end{equation}
where $A_{\rm sys}$ is the amplitude of the correlation.  
An additional constant offset was also fit but was consistent with zero and is therefore omitted from Equation \ref{eqn:sysfit}.  We
obtain $A_{\rm sys} = 0.56\pm0.11$~$\mu$as$^2$~yr$^{-2}$, which is significant and presents a prohibitive
obstacle to measuring the astrometric HD signal using this uncertainty-based selection method.

\begin{figure}[t]
\begin{centering}
  \includegraphics[scale=0.25,trim= 20 40 20 20,clip=false]{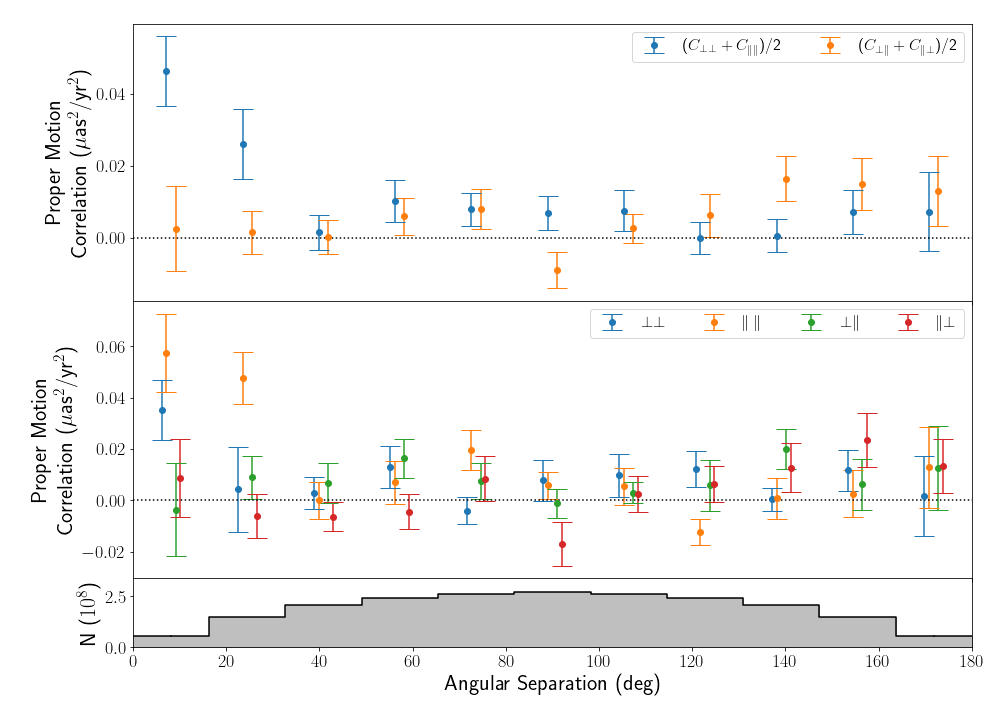}
  \caption{ Angular correlations of quasar proper motions show
    significant excess power when the E- and B-mode dipoles are not removed.
    Top:  aligned parallel and perpendicular modes (blue) and mixed modes (orange) vs.\ the angular separation of quasar pairs for 64,879 quasars selected for proper motion amplitude $< 100$~$\mu$as~yr$^{-1}$.   The points are laterally offset from the center of each angular separation bin for clarity.
    Middle:  all four individual proper motion correlation combinations.  
    Bottom:  the distribution of quasar pairs, scaled by $10^8$.}
\label{fig:noE1B1sub}
\end{centering}
\end{figure}

For the 64,879-quasar proper motion-selected sample,
the number of unique pairs used for the astrometric HD calculation is 2,104,609,881.
In this sample, the redshift quartiles are $z=0.93$, 1.41, and 1.92, corresponding to lookback times
of 7.4, 9.0, and 10.1 Gyr and minimum GW frequencies of $4.3\times10^{-18}$, $3.5\times10^{-18}$, and
$3.1\times10^{-18}$ Hz.  
We fit and subtract the E- and B-mode proper motion dipoles (using VSH fitting) from the proper motions of
this specific sample in order to isolate higher modes.  When these are not removed from the proper motion catalog,
significant excess power remains and the astrometric HD signal cannot be recovered or constrained.
Figure \ref{fig:noE1B1sub} shows the impact of failing to remove the dipoles before calculating the proper motion
angular correlations.  Most notably, the aligned proper motion correlations show a significant
3.9$\sigma$ positive error-weighted mean.

In order to obtain the proper motion projections used in the astrometric HD calculation we need to
find the great-circle directions connecting an arbitrary pair of quasars.  
First, we construct the radial unit vector corresponding to the equatorial coordinates $(\alpha_n,\delta_n)$ of object $n$:
\begin{equation}
  \bm{\hat{n}} = \left(
    \begin{array}{c}
      \cos\alpha_n \cos\delta_n \\
      \sin \alpha_n \cos\delta_n \\
      \sin\delta_n
    \end{array} \right).
\end{equation}
For a second object in direction $\bm{\hat{m}}$, the angular separation $\theta$ between $\bm{\hat{n}}$ and $\bm{\hat{m}}$ is
\begin{equation}
  \cos\theta = \bm{\hat{n}}\cdot\bm{\hat{m}}
\end{equation}
This angle $\theta$ is the abscissa in the HD curve (Equation \ref{eqn:gamma} and Figure \ref{fig:theory}).  

The 3D proper motion vector is the time derivative of the position unit vector:
\begin{equation}
  \bm\mu_n  =  \left(
    \begin{array}{c}
      -  \mu_{n,\alpha} \sin\alpha_n \cos\delta_n  - \mu_{n,\delta}  \cos\alpha_n \sin\delta_n \\
       \mu_{n,\alpha} \cos \alpha_n \cos \delta_n - \mu_{n,\delta} \sin \alpha_n \sin \delta_n  \\
      \mu_{n,\delta} \cos \delta_n 
\end{array}     \right) ,
\end{equation}
Note that $\bm\mu_n$ lies on the celestial sphere, as required:  $\bm{\hat{n}} \cdot \bm\mu_n = 0$.  Also, all of the time derivatives are
with respect to the present cosmic time, $t_0$, and we neglect the minuscule radial change in the quasar position due to peculiar motions, the secular redshift drift \citep{darling2012,moresco2022}, and the radial component of gravitational waves.  The latter can be cross-correlated with
angular motions, and will likely be a powerful tool for combined pulsar timing and astrometry \citep[e.g.,][]{mihaylov2018}.

The proper motion of each quasar in direction $\bm{\hat{n}}$ will have a component along and perpendicular to the great circle
path to the quasar at $\bm{\hat{m}}$, and the same will be true for the proper motion of $\bm{\hat{m}}$ with respect to the path to $\bm{\hat{n}}$.  
The unit vector $\bm{\hat{e}}_\perp$ perpendicular to the great circle connecting $\bm{\hat{n}}$ to $\bm{\hat{m}}$ is the same for $n$ and $m$ but the unit vectors along
the great circle are not the same and will be designated $\bm{\hat{e}}_{\parallel,n}$ and $\bm{\hat{e}}_{\parallel,m}$.
We construct these unit vectors perpendicular and parallel to the great circle that connects $\bm{\hat{n}}$ to $\bm{\hat{m}}$:
\begin{eqnarray}
  \bm{\hat{e}}_\perp &=& {\bm{\hat{n}}\times\bm{\hat{m}}\over \sqrt{1-(\bm{\hat{n}}\cdot \bm{\hat{m}})^2}}\\
  \bm{\hat{e}}_{\parallel,n} &=& \bm{\hat{e}}_\perp \times \bm{\hat{n}} \\
  \bm{\hat{e}}_{\parallel,m} &=& \bm{\hat{e}}_\perp \times \bm{\hat{m}}
\end{eqnarray}
\citep{mihaylov2018,caliskan2024}.  

We calculate four proper motion correlations between pairs of quasars, corresponding to all combinations of
motion perpendicular to and parallel to the great circle connecting each pair.  The two correlations that
are sensitive to a stochastic gravitational wave background are:
\begin{eqnarray} \label{eqn:Cperpperp}
  C_{\perp\perp}(\theta)  &\equiv&  \left\langle \mu_{n,\perp}\ \mu_{m,\perp} \right\rangle_\theta
      = \left\langle (\bm\mu_n \cdot \bm{\hat{e}}_\perp ) (\bm\mu_m \cdot \bm{\hat{e}}_\perp )\right\rangle_\theta\\
  C_{\parallel\parallel}(\theta)  &\equiv&  \left\langle \mu_{n,\parallel}\ \mu_{m,\parallel} \right\rangle_\theta = 
      \left\langle (\bm\mu_n \cdot \bm{\hat{e}}_{\parallel,n} ) (\bm\mu_m \cdot \bm{\hat{e}}_{\parallel,m} )\right\rangle_\theta .\label{eqn:Cparpar}
\end{eqnarray}
The angle brackets indicate an average of all pairs at a given $\theta$ (in practice one uses a range $\Delta\theta$ centered at $\theta$).  
The two cross-projection correlations are
\begin{eqnarray}\label{eqn:Cperppar}
  C_{\perp\parallel}(\theta) & \equiv  \left\langle \mu_{n,\perp} \mu_{m,\parallel} \right\rangle_\theta & =  \left\langle (\bm\mu_n \cdot \bm{\hat{e}}_\perp ) (\bm\mu_m \cdot \bm{\hat{e}}_{\parallel,m} )\right\rangle_\theta\\
  C_{\parallel\perp}(\theta) & \equiv  \left\langle \mu_{n,\parallel} \mu_{m,\perp} \right\rangle_\theta & =  \left\langle (\bm\mu_n \cdot \bm{\hat{e}}_{\parallel,n} ) (\bm\mu_m \cdot \bm{\hat{e}}_\perp )\right\rangle_\theta.\label{eqn:Cparperp}
\end{eqnarray}
These should be zero for unpolarized transverse traceless gravitational waves \citep[e.g.,][]{qin2019}.  
These cross-term correlations provide a powerful check on systematics and can be compared in a one-to-one fashion to
the aligned-projection correlations.   Absent systematic structure, the cross terms also assess the statistical errors:  comparison of
the scatter to the uncertainties can be used to assess the appropriateness of the uncertainty estimates and the independence of the measurements.

To calculate the four correlations, we project the proper motions for each quasar pair parallel and perpendicular to each pair's
great circle, form the product of these projections and their cross-terms, bin in angular separation $\theta$, and compute an error-weighted mean of each angular separation bin.  To estimate the uncertainties in each bin, we bootstrap the error-weighted mean based on resampling quasar pairs.  The bootstrap reveals no bias in the error-weighted mean.
The number of angular separation bins is chosen to maximize signal-to-noise while adequately sampling the astrometric
HD curve.  We deliberately use an odd number of bins in order to sample any signal at $\theta = 90^\circ$, either in
the aligned or cross terms.  

%
%

\begin{figure*}[t]
\begin{centering}
  \includegraphics[scale=0.5,trim= 20 20 20 20,clip=false]{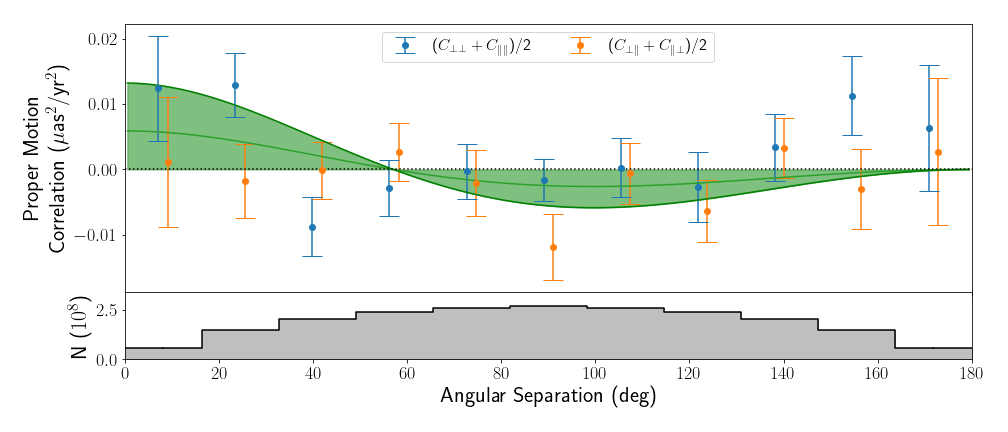}
  \caption{Proper motion power in parallel and perpendicular modes (blue) and mixed modes (orange) vs.\ the angular separation of quasar pairs for 64,879 quasars selected based on their proper motion amplitude $< 100$~$\mu$as~yr$^{-1}$.   The points are offset by $\pm1^\circ$ from the center of each angular separation bin for clarity.  
    The green line shows the best-fit (and non-significant amplitude) astrometric HD curve to the aligned correlations,  $(C_{\perp\perp}+C_{\parallel\parallel})/2$, and the filled region shows the $< 95$\% confidence envelope.
    This envelope equates to 95\% confidence limits on the characteristic strain
    $h_{c} \leq 2.7 \times 10^{-12}$ and the gravitational wave energy density $h_{70}^2\Omega_{GW} \leq 0.0096$ at reference frequency $f_{\rm ref} = 1$~yr$^{-1}$.
    The mixed mode measurements should be consistent with zero and provide a control on systematics.
    The lower histogram shows the distribution of quasar pairs, scaled by $10^8$.}
\label{fig:HDcurve}
\end{centering}
\end{figure*}

\section{\label{sec:results} Results}

\subsection{Vector Spherical Harmonics}\label{subsec:VSHresults}

\begin{deluxetable}{cr|cr}
\tabletypesize{\scriptsize} 
\tablecaption{Quaia 20.5 Dipole and Quadrupole VSH Fits \label{tab:vsh}}
\tablewidth{0pt} 
\tablehead{
\colhead{Parameter} & \colhead{Value} &
\colhead{Parameter} & \colhead{Value} \\
\colhead{} & \colhead{($\mu$as yr$^{-1}$)} & 
\colhead{} & \colhead{($\mu$as yr$^{-1}$)}}
\startdata 
\multicolumn{4}{c}{\bf \bm{$E$}-Modes}\\
\multicolumn{2}{l}{Dipole (Aberration Drift)} & Quadrupole (GWs)\\
$s_{10}$       & $-4.06(0.90)$ & $s_{20}$       & $-3.70(0.87)$\\   
$s_{11}^{Re}$ & $+0.67(0.66)$ & $s_{21}^{Re}$ & $+3.72(0.62)$ \\
$s_{11}^{Im}$ & $-10.22(0.63)$ & $s_{21}^{Im}$ &$+2.47(0.65)$ \\
\bm{$\sqrt{P_1^s}$} & \bm{$15.04(0.89)$} & $s_{22}^{Re}$ &$-2.76(0.68)$ \\
{\bf Amplitude} & \bm{$5.20(0.31)$} & $s_{22}^{Im}$ & $+2.34(0.68)$ \\
Z-score & 13.1 & \bm{$\sqrt{P_2^s}$} & \bm{$8.92(0.91)$} \\
\hline
\noalign{\vskip 1mm} 
\multicolumn{4}{c}{\bf \bm{$B$}-Modes}\\
\multicolumn{2}{l}{Dipole (Rotation)} & \multicolumn{2}{l}{Quadrupole (GWs)}\\
$t_{10}$       & $-4.08(0.99)$ & $t_{20}$ & $+1.80(0.85)$ \\
$t_{11}^{Re}$ & $-5.03(0.66)$ &  $t_{21}^{Re}$ & $+0.89(0.70)$\\
$t_{11}^{Im}$ & $-2.66(0.60)$ &  $t_{21}^{Im}$ & $+1.62(0.68)$\\
\bm{$\sqrt{P_1^t}$} & \bm{$9.02(0.93)$} & $t_{22}^{Re}$ & $-1.31(0.63)$\\
{\bf Amplitude} & \bm{$3.12(0.32)$} & $t_{22}^{Im}$ & $+0.91(0.63)$ \\
Z-score & 8.2 & \bm{$\sqrt{P_2^t}$} & \bm{$3.89(0.92)$}\\
\hline
\noalign{\vskip 1mm} 
\multicolumn{4}{c}{\bf Combined Modes}\\
\bm{$\sqrt{P_1}$} & \bm{$17.54(0.90)$} &  \bm{$\sqrt{P_2}$} & \bm{$9.74(0.91)$} \\
Z-score & 15.5 &  Z-score & 8.5
\enddata 
\tablecomments{Proper motions are simultaneously fit to electric and magnetic dipole vector fields, the dipoles are subtracted
  from the proper motion data, and then quadrupoles are simultaneously fit to the residual.  The VSH coefficients, power, and amplitudes are defined in Appendix \ref{sec:vsheqns}.  
Parenthetical quantities indicate 1$\sigma$ uncertainties.  The Z-score statistic, defined in \citet{mignard2012}, is unitless.}
\end{deluxetable}

Table \ref{tab:vsh} shows the best-fit VSH dipole and quadrupole coefficients, power, and amplitudes (defined in Appendix
\ref{sec:vsheqns}).
The VSH fits show significant power in the quadrupole with unequal contributions from
the E2 and B2 modes, which suggests that there are significant systematic effects in the proper motion vector field.
The E2 mode is significant, but the B2 mode is not.
If we nonetheless use the total quadrupole power to estimate the energy density in gravitational waves,
we obtain $\Omega_{GW} = 0.042(0.008)$.  This is consistent with the non-significant 
\citet{jaraba2023} value of $h_{70}^2\Omega_{GW} = 0.040(0.017)$ for their ``Intersection'' sample of 773,471 quasars
with very similar frequency sensitivity.
If we assume that the non-significant B2 mode is the most constraining and that the significant E2 power is not physical,
then we can use the B2 mode as the statistical floor on half of the total power, set both modes to
be equal to $P_2^t$, and obtain the non-significant value of  $\Omega_{GW} = 0.013(0.006)$ or $h_{70}^2\Omega_{GW} \leq 0.023$ at 95\% confidence, which is nearly identical to VLBI measurements \citep{jaraba2023}.  This is a new sensitivity
limit for optical astrometry and the first time that it is similar to VLBI.  This is also a three-fold improvement over recent work using Gaia DR3 quasars.

\subsection{Astrometric Hellings-Downs Curve}

Figure \ref{fig:HDcurve} shows the astrometric HD curve for the aligned and mixed correlations.  The best-measured
angular scales are roughly between 30$^\circ$ and 150$^\circ$ where the number of pairs is $>2\times10^8$ and
the uncertainties are remarkably small, $\sim$0.005 $\mu$as$^2$ yr$^{-2}$.
The uncertainties grow by a factor of two for close pairs and antipodal pairs where the number of pairs drops to $6\times10^7$.  There is no significant
power in the astrometric correlations, either point-by-point, in the error-weighted mean power, or based on fitting the HD curve to the aligned correlations data (see below).  The scatter in the data is consistent with the uncertainty estimates.
The error-weighted mean power in the average of the aligned correlations is $0.0009(0.0015)$~$\mu$as$^2$~yr$^{-2}$.
The error-weighted mean is $-0.0017(0.0016)$~$\mu$as$^2$~yr$^{-2}$ for the cross-terms.  
The noise-consistent measurements and the consistency between the correlations that should and should not show
a gravitational wave signal demonstrate that we have successfully removed proper motion systematics but have not
yet produced an astrometric detection of gravitational waves.  There is
also little or no evidence of a noise floor:  as we grew the sample size, the uncertainties in the HD curve decreased correspondingly and as expected.  

A fit of the astrometric HD curve to the aligned correlations, Equations \ref{eqn:gamma} and \ref{eqn:CtoIGamma},
produces a non-significant amplitude:  $I_{GW} f^3 = 0.0029(0.0024)$~$\mu$as$^2$ yr$^{-2}$.  A direct fit to the
  proper motions does not significantly differ from the fit to the binned data.
For a 95\% confidence limit of $f^3 I_{GW}(f) \leq 0.0063$~$\mu$as$^2$ yr$^{-2}$,
we obtain a limit on the characteristic strain of
\begin{eqnarray}
  h_{c} \leq 2.7 \times 10^{-12} \nonumber
\end{eqnarray}
for $f_{\rm ref} = 1$~yr$^{-1}$.  The corresponding limit on the gravitational wave energy density is 
\begin{eqnarray}
  h_{70}^2\Omega_{GW} \leq 0.0096 \nonumber
\end{eqnarray}
also at 95\% confidence.  This is better than a factor of two lower than the VSH measurement
corrected for non-equal E- and B-mode quadrupoles and is a factor of 4.4 times lower than the
uncorrected quadrupole measurement.  This is the strongest astrometric constraint on gravitational waves to date.

%
%

\section{\label{sec:discussion}Discussion and Conclusions}

Are these new astrometric limits on strain and energy density constraining on possible sources of the pulsar-timing-based
stochastic gravitational wave background?
The characteristic gravitational wave strain claimed by pulsar timing is $h_c = 2.4^{+0.7}_{-0.6}\times10^{-15}$ at
a reference frequency of $f_{\rm ref} = 1$~yr$^{-1}$, assuming the strain spectrum appropriate for an ensemble
of massive black hole binaries, $h_c(f) \propto f^{-2/3}$ \citep{agazie2023}.  This spectral index and strain measurement
predict a strain of $2.4\times10^{-14}$ at 1 nHz (1~yr$^{-1}$ is 31.7 nHz), roughly the expected turnover frequency for massive black hole binaries.  This is well below the limits obtained from VSH fitting and the astrometric HD curve for quasars, so the
astrometry is not constraining on extant observations.  It may constrain exotic interpretations of the gravitational wave background.

This work, however, has produced the first astrometric HD curves and demonstrated a powerful approach to the two
transverse dimensions of gravitational wave detection not probed by pulsar timing.  We have also shown how the systematics present
in Gaia VSH analysis can be sidestepped and quantified using pairwise angular motions.  We have produced the best limits on
the stochastic astrometric gravitational wave background to date, including demonstrating for the first time that optical
astrometry can surpass the limits from VLBI, mostly due to sample size.  

Future Gaia data releases will improve this measurement, particularly due to longer time baselines (astrometric
uncertainties scale roughly as $t^{-3/2}$ if sampling is regular), but the sample size may grow as well.  We also expect
cross-correlations between quasar proper motions and pulsar timing to produce interesting limits or detections.  This
cross-correlation will sample the full three-dimensional realm of observable gravitational wave effects and can probe the polarization, isotropy, and transverse traceless nature of gravitational waves.  

\acknowledgments
We thank A. Williams, D. Hogg, and K. Storey-Fisher for discussion of the Quaia catalog and K. Inomata for helpful discussion
of astrometric correlations.  We also thank the referee for valuable feedback.
  This research made use of NumPy \citep{NumPy}, Matplotlib \citep{Matplotlib}, and
  Astropy\footnote{\url{http://www.astropy.org}}, a community-developed core Python package for Astronomy \citep{astropy:2013, astropy:2018}.

  \facility{Gaia}
  \software{astropy \citep{astropy:2013, astropy:2018}, NumPy \citep{NumPy}, Matplotlib \citep{Matplotlib},
    \texttt{lmfit} \citep{newville2021}, \texttt{emcee} \citep{emcee}, TOPCAT \citep{taylor2005}.}

\bibliography{ms}

\appendix

\section{Dipole and Quadrupole Vector Spherical Harmonics}\label{sec:vsheqns}

The VSH modes that are fit to vector fields are obtained from the divergence and curl of
the scalar spherical harmonics $Y_{\ell m}$:
\begin{equation}
\mathbf{S}_{\ell m}(\alpha,\delta) = {1 \over \sqrt{\ell(\ell+1)}}\ \bm{\nabla} Y_{\ell m}(\alpha,\delta)
\end{equation}
and
\begin{equation}
\mathbf{T}_{\ell m}(\alpha,\delta) = {-1 \over \sqrt{\ell(\ell+1)}}\ \hat{n} \times \bm{\nabla} Y_{\ell m}(\alpha,\delta)
\end{equation}
where the $\mathbf{S}_{\ell m}$ is the ``spheroidal'' curl-free $E$-mode of degree $\ell$ and order $m$, 
$\mathbf{T}_{\ell m}$ is the ``toroidal'' divergence-less $B$-mode, and $\hat{n}$ is the radial unit vector \citep{mignard2012}.  
The E-modes, B-modes, and radial vector $\hat{n}$ are mutually orthogonal by construction.
A general vector field on a sphere can be fully described by a sum of these VSH with complex
spheroidal and toroidal coefficients $s_{\ell m}$ and $t_{\ell m}$:
\begin{equation}\label{eqn:vsh}
  \mathbf{V}(\alpha,\delta) = \sum_{\ell=1}^{\infty} \sum_{m=-\ell}^\ell
                   (s_{\ell m} \mathbf{S}_{\ell m}(\alpha,\delta) + t_{\ell m} \mathbf{T}_{\ell m}(\alpha,\delta) ).
\end{equation}
The decomposition of a real vector field into VSH always produces real-valued solutions.

Following the formalism of \citet{mignard2012}, the dipole vector fields $\mathbf{V}_{E1}$
and $\mathbf{V}_{B1}$ have the form
\begin{eqnarray}
  \mathbf{V}_{E1} (\alpha,\delta) & = &
      {1\over2} \sqrt{3\over\pi}\, \left(s_{11}^{Re}\, \, \sin\alpha+s_{11}^{Im}\,  \cos\alpha\right) \mathbf{\hat{e}}_\alpha 
                                         +{1\over2} \sqrt{3\over\pi}\, \left(s_{10}\, \sqrt{1\over 2}\, \cos\delta + s_{11}^{Re}\, \cos\alpha\sin\delta
       -s_{11}^{Im}\,  \sin\alpha\sin\delta\right) \mathbf{\hat{e}}_\delta \\
\mathbf{V}_{B1} (\alpha,\delta) & = &
     {1\over2} \sqrt{3\over\pi}\, \left(t_{10}\, \sqrt{1\over 2}\, \cos\delta + t_{11}^{Re}\,  \cos\alpha\sin\delta
       -t_{11}^{Im}\,  \sin\alpha\sin\delta\right) \mathbf{\hat{e}}_\alpha 
                                         + {1\over2} \sqrt{3\over\pi}\,\left(-t_{11}^{Re}\,  \sin\alpha-t_{11}^{Im}\, \cos\alpha\right) \mathbf{\hat{e}}_\delta 
\end{eqnarray}
\noindent
where
$\mathbf{\hat{e}}_\alpha$ and $\mathbf{\hat{e}}_\delta$ are the unit vectors in the R.A. and declination directions, respectively.
The superscripts $Re$ and $Im$ indicate the real or imaginary parts of the coefficients (the $m=0$ coefficients are always real). 

The maximum dipole proper motion, which we call its amplitude, is calculated via
\begin{equation}
  A_{\rm E1, B1} = {\sqrt{P^{s,t}_\ell} \over 2\sqrt{2 \pi/3}}
  \end{equation}
and the dipole apex directions $\alpha_{\rm E1, B1}^*$ and $\delta_{\rm E1, B1}^*$ are
\begin{eqnarray}
  \tan\alpha_{\rm E1}^* &=& -s_{11}^{Im}\over s_{11}^{Re} \\
  \tan\delta_{\rm E1}^* &=&  {-s_{10} \over \sqrt{2} \left(s_{11}^{Re} \cos \alpha_{\rm E1}^* - s_{11}^{Im}\sin\alpha_{\rm E1}^*\right)} \\
  \tan\alpha_{\rm B1}^* &=& -t_{11}^{Im}\over t_{11}^{Re} \\
  \tan\delta_{\rm B1}^* &=&  {-t_{10} \over \sqrt{2} \left(t_{11}^{Re} \cos \alpha_{\rm B1}^* - t_{11}^{Im}\sin\alpha_{\rm B1}^*\right)}.
\end{eqnarray}

The quadrupole vector fields are
\begin{eqnarray}
  \mathbf{V}_{E2} (\alpha,\delta) &=& 
     {1\over2} \sqrt{5\over\pi}\, \left(s_{21}^{Re}\,  \sin\alpha\sin\delta+s_{21}^{Im}\,  \cos\alpha\sin\delta
- s_{22}^{Re}\,  \sin2\alpha\cos\delta \right.
       \left. -s_{22}^{Im}\,  \cos2\alpha\cos\delta\right) \mathbf{\hat{e}}_\alpha \nonumber\\
     &+& {1\over2} \sqrt{5\over\pi}\, \left(s_{20}\, {1\over2}\sqrt{3\over 2}\, \sin2\delta - s_{21}^{Re}\,  \cos\alpha\cos2\delta
     + s_{21}^{Im}\,  \sin\alpha\cos2\delta \right.
        - s_{22}^{Re}\, {1\over2}\, \cos2\alpha\sin2\delta \nonumber\\
      & &  \ \ + \left. s_{22}^{Im}\, {1\over2}\, \sin2\alpha\sin2\delta \right) \mathbf{\hat{e}}_\delta  \\
  \label{eqn:Equad}
  \mathbf{V}_{B2} (\alpha,\delta) &=& 
     {1\over2} \sqrt{5\over\pi}\, \left(t_{20}\, {1\over2}\sqrt{3\over 2}\, \sin2\delta - t_{21}^{Re}\,  \cos\alpha\cos2\delta
     + t_{21}^{Im}\,  \sin\alpha\cos2\delta \right.
        - t_{22}^{Re}\, {1\over2}\, \cos2\alpha\sin2\delta \nonumber\\
      & & \ \ + \left.t_{22}^{Im}\, {1\over2}\, \sin2\alpha\sin2\delta \right) \mathbf{\hat{e}}_\alpha \nonumber\\
      &+&{1\over2} \sqrt{5\over\pi}\, \left(-t_{21}^{Re}\,  \sin\alpha\sin\delta-t_{21}^{Im}\,  \cos\alpha\sin\delta
      + t_{22}^{Re}\,  \sin2\alpha\cos\delta 
       + t_{22}^{Im}\,  \cos2\alpha\cos\delta\right) \mathbf{\hat{e}}_\delta .
\end{eqnarray}

The power in mode $\ell$ is obtained from the quadrature sum of coefficients (modulo factors of 2):
\begin{equation}\label{eqn:power}
  P_\ell = s_{\ell 0}^2 + t_{\ell 0}^2 +2 \sum_{m=1}^{\ell} \left[ (s_{\ell m}^{Re})^2+ (s_{\ell m}^{Im})^2+ (t_{\ell m}^{Re})^2+ (t_{\ell m}^{Im})^2\right]
\end{equation}
\citep{mignard2012}.  To compute the power in E$\ell$ or B$\ell$ modes separately, simply omit the $t_{\ell m}$ or
$s_{\ell m}$ terms in the sum.

\end{document}